\documentclass[preprint,review,12pt]{elsarticle}
\usepackage{amssymb}
\usepackage{hyperref}
\usepackage{amsmath}
\usepackage{cleveref}
\usepackage{booktabs}
\usepackage{caption,setspace}
\usepackage{tabularx}
\usepackage[export]{adjustbox}
\usepackage{tikz}
\usepackage{geometry}
\usepackage{wrapfig} % biography
\usepackage{flushend} % align
\usepackage{algorithm}
\usepackage{algpseudocode}
\usepackage{xcolor}
\usepackage{multirow}

\hypersetup{pdfauthor={Name}}

\makeatletter
\newcommand{\ssymbol}[1]{^{\@fnsymbol{#1}}}
\makeatother

\usepackage{array}
\newcolumntype{P}[1]{>{\centering\arraybackslash}p{#1}}

\AtBeginDocument{
    \crefformat{equation}{#2Eq.~#1#3}
    \crefformat{figure}{#2Fig.~#1#3}
    \crefformat{table}{#2Tab.~#1#3}
}

\journal{Knowledge-Based Systems} % arXiv

\begin{document}
\begin{sloppypar}

\begin{frontmatter}

\title{MedAugment: Universal Automatic Data Augmentation Plug-in for Medical Image Analysis}

%% ---------------------------------------- With Aff ---------------------------------------- %%

\author[a]{Zhaoshan Liu\corref{cor1}}
\ead{e0575844@u.nus.edu}

\author[b]{Qiujie Lv\corref{cor1}}
\ead{lvqiujie@zzu.edu.cn} 

\author[a]{Yifan Li}
\ead{e0576095@u.nus.edu} 

\author[c]{Ziduo Yang}
\ead{yangzd@jnu.edu.cn} 

\author[a]{Lei Shen\corref{cor2}}
\ead{mpeshel@nus.edu.sg}

\cortext[cor1]{Equal contribution}
\cortext[cor2]{Corresponding author}

\address[a]{Department of Mechanical Engineering, National University of Singapore, 9 Engineering Drive 1, Singapore, 117575, Singapore}

\address[b]{School of Computer and Artificial Intelligence, Zhengzhou University, 100 Science Avenue, Zhengzhou, 450001, China}

\address[c]{Department of Electronic Engineering, Jinan University, 601 West Huangpu Avenue, Guangzhou, 510632, China}

%% ---------------------------------------- Main ---------------------------------------- %%

\begin{abstract}
Data augmentation (DA) has been widely leveraged in computer vision to alleviate data shortage, while its application in medical imaging faces multiple challenges. The prevalent DA approaches in medical image analysis encompass conventional DA, synthetic DA, and automatic DA. However, these approaches may result in experience-driven design and intensive computation costs. Here, we propose a suitable yet general automatic DA method for medical images termed MedAugment. We propose pixel and spatial augmentation spaces and exclude the operations that can break medical details and features. Besides, we propose a sampling strategy by sampling a limited number of operations from the two spaces. Moreover, we present a hyperparameter mapping relationship to produce a rational augmentation level and make the MedAugment fully controllable using a single hyperparameter. These configurations settle the differences between natural and medical images. Extensive experimental results on four classification and four segmentation datasets demonstrate the superiority of MedAugment. Compared with existing approaches, the proposed MedAugment prevents producing color distortions or structural alterations while involving negligible computational overhead. Our method can serve as a plugin without an extra training stage, offering significant benefits to the community and medical experts lacking a deep learning foundation. The code is available at \href{https://github.com/NUS-Tim/MedAugment}{https://github.com/NUS-Tim/MedAugment}.
\end{abstract}

%% Graphical abstract
% \begin{graphicalabstract}
% \includegraphics{grabs}
% \end{graphicalabstract}

\begin{keyword}
%% keywords here, in the form: keyword \sep keyword
Data Augmentation \sep Medical Image Analysis \sep Image Classification \sep Image Segmentation

%% PACS codes here, in the form: \PACS code \sep code

%% MSC codes here, in the form: \MSC code \sep code
%% or \MSC[2008] code \sep code (2000 is the default)

\end{keyword}

\end{frontmatter}

%% \linenumbers

%% ---------------------------------------- Introduction ---------------------------------------- %%

\section{Introduction}
\label{1}

Medical image analysis (MIA) employs various imaging modalities to visually create an interior body representation and assist with further medical diagnoses. Currently, MIA is predominantly conducted by medical experts, and this time-consuming and labor-intensive process can potentially result in variability in interpretation and accuracy. To this end, deep learning (DL) \citep{yan2021precise, krizhevsky2012imagenet, yan2022age} has been adopted into the MIA field for assistance, especially for mainstream classification and segmentation tasks. Though DL-based MIA has achieved promising results \citep{poonkodi20233d, chen2022uncertainty, li2023transforming}, ensuring the performance of the DL model under data scarcity can be challenging. Differing from natural images, the scarcity of data in MIA can be attributed to two primary factors. Firstly, collecting medical images necessitates specialized equipment and requires expert annotation. Secondly, the distribution of collected images is constrained by patient privacy concerns \citep{liu2023gsda}. In this context, various techniques have been proposed to mitigate the data shortage, and data augmentation (DA) is the most prevalent and effective one \citep{eisenmann2023winner, ouyang2022causality}. The DA improves the performance and generalization capability of the model by enhancing the diversity and richness of data, and its prevalent usage in the realm of MIA includes conventional DA, synthetic DA, and automatic DA.

The conventional DA is one of the most common DA approaches \citep{khened2019fully, kaushik2021diabetic, isensee2021nnu}. It consists of various DA operations such as rotation, flip, and translation \citep{eisenmann2023winner} to compose varying DA pipelines. Though these methods are straightforward and effective, the pipeline design, such as operation selection, sequence adjustment, and magnitude determination, heavily relies on experience. This makes conventional DA unsuitable for personnel without a solid DL foundation and can lead to suboptimal augmentation diversity. Leveraging generative adversarial network (GAN) \citep{chai2022synthetic, li2021semantic, iqbal2023unet, pang2021semi, beers2018high} is one of the most prevalent synthetic DA methods. The GAN encompasses the generator and discriminator playing an adversarial game, and is capable of synthesizing results at the pixel level. However, GAN-based approaches are time-consuming, data-hungry \citep{feng2021gans, karras2019style}, and can produce varied synthesized quality. Compared to GAN, diffusion model \citep{moghadam2023morphology, pinaya2022brain, khader2023denoising, tang2023multi} is a synthetic alternative. However, its applications face low sampling speed and high computational cost \citep{khader2023denoising}. The performance of automatic DA \citep{qin2020automatic, xu2020automatic, lyu2022aadg} has been recently well-proved. The automatic DA consists of an augmentation space with conventional operations, and the input is augmented through varying operations sampled from the space \citep{cubuk2019autoaugment, cubuk2020randaugment, muller2021trivialaugment, lingchen2020uniformaugment}. Though these methods can increase data diversity and richness, they either introduce additional overhead or lack adaptation to medical imaging.

To tackle data shortage \citep{pang2021semi, liu2023recent, ouyang2022self} and augmentation challenges encountered, we propose a suitable yet general automatic DA method MedAugment for MIA. Compared to existing methods with a single augmentation space, we present two augmentation spaces termed pixel augmentation space $A_{p}$ and spatial augmentation space $A_{s}$ and exclude the operations that can disrupt the details and features in medical images. This can prevent severe color distortions or structural alterations and ensure the diagnostic value. Besides, we propose an operation sampling strategy by constraining the number of operations sampled from the two spaces. Moreover, we present a hyperparameter mapping relationship to produce a rational augmentation level and make the MedAugment fully controllable with a single hyperparameter. These designs effectively tackle the differences between natural and medical images. Extensive experimental results on four classification and four segmentation datasets demonstrate the leadership of the proposed MedAugment. Compared with existing methods, the proposed method prevents color distortions or structural alterations while introducing negligible computational overhead. The MedAugment can serve as a plugin without any extra training stage, benefiting the MIA community and medical experts without a solid foundation in DL. To sum up, our main contributions are:
\begin{itemize}
    \item We propose a suitable yet general automatic data augmentation method termed MedAugment for medical image analysis.
    \item We present pixel and spatial augmentation spaces, a sampling strategy, and a hyperparameter mapping relationship.
    \item We perform comprehensive experiments on eight datasets, and the results demonstrate the superiority of the MedAugment.
\end{itemize}
 
The rest of this paper is organized as follows. Section \hyperref[2]{2} "Related Work" illustrates the recent progress in automatic DA and DA in MIA. Section \hyperref[3]{3} "Methods" discusses the methodology of the proposed MedAugment. In Section \hyperref[4]{4} "Experiments", the datasets leveraged and experimental setup are introduced. We illustrate the results, analysis, and ablation study in Section \hyperref[5]{5} "Results and Analysis". We summarize our work and point out the future perspectives in Section \hyperref[6]{6} "Conclusions".

%% ---------------------------------------- Related work ---------------------------------------- %%

\section{Related Work}
\label{2}

\subsection{Automatic Data Augmentation}
\label{2.1}

Numerous automatic DA methods have been developed to combine conventional operations. In 2019, Cubuk et al. \cite{cubuk2019autoaugment} developed an AutoAugment where a policy in the search space is composed of several sub-policies, and each sub-policy is randomly selected for each image. Each sub-policy consists of two DA operations selected from sixteen. Though AutoAugment achieves promising performance, the DA policy is searched using the reinforcement learning method and thus can be computationally expensive. To this end, Lim et al. \citep{lim2019fast} proposed the Fast AutoAugment to identify the augmentation policy by employing density matching across paired training datasets. The Fast AutoAugment is based on Bayesian DA \citep{tran2017bayesian} and can recover additional missing data points through Bayesian optimization during the policy search phase. Ho et al. \citep{ho2019population} presented a population-based augmentation approach to produce the nonstationary policy rather than the fixed one. Although these approaches effectively reduce the search cost, a distinct search phase remains necessary. Zhang et al. \citep{zhang2019adversarial} introduced an adversarial autoaugment approach, which can simultaneously optimize the target model and the augmentation policy search loss. Li et al. \citep{li2020differentiable} proposed a differentiable automatic DA method to relax the discrete DA policy selection to a differentiable optimization problem via Gumbel-Softmax. These methods shift policy search from an explicit, separate stage to an implicit, training-time optimization process, while policy optimization remains involved.

To eliminate policy search, Cubuk et al. developed a RandAugment \citep{cubuk2020randaugment} method, in which multiple DA operations with the same augmentation level are sequentially leveraged. The augmentation space of RandAugment comprises fourteen operations. Comparable work of RandAugment includes the TrivialAugment \citep{muller2021trivialaugment} that utilizes a single operation and samples the augmentation level anew for each image. Besides, the UniformAugment \citep{lingchen2020uniformaugment} fixes the number of operations to two and drops each operation with a probability $p = 0.5$. Besides leveraging DA operations successively, an alternative is to combine them in parallel. For instance, the AugMix \citep{hendrycks2019augmix} randomly samples several operations from nine to compose an augmentation chain. Several augmentation chains and a separate chain without DA are mixed based on their weights to derive the augmented images. Though these approaches are effective and low-computation, their usage poses various challenges in MIA. Firstly, the involved operations, such as \texttt{invert}, \texttt{equalize}, and \texttt{solarize}, can disrupt the intricate details and features characteristic of medical images. Secondly, the sampling strategy tends to overlook the fact that medical images exhibit heightened sensitivity to operations such as \texttt{brightness}, \texttt{contrast}, and \texttt{posterize}. Finally, image mixing presents challenges in processing masks, limiting the application in medical segmentation.

\subsection{Data Augmentation in MIA}
\label{2.2}

A large proportion of studies leverage conventional DA. For example, Kaushik et al. \citep{kaushik2021diabetic} utilized translation, rotation, scale, flip, etc., to augment fundus images for diabetic retinopathy diagnosis. Khened et al. \citep{khened2019fully} augmented the dataset using rotation, translation, scale, Gaussian noise, etc., for cardiac segmentation. Zhang et al. \citep{zhang2020generalizing} proposed a BigAug approach, which utilizes a stacked transformation sequence to generalize segmentation models to unseen domains. The DA transformations employed primarily alter image quality, appearance, and spatial configuration. Chen et al. \citep{chen2022enhancing} introduced an AdvChain approach to optimize the DA transformation parameters by simultaneously considering visual information and network fragility. Besides, Isensee et al. \citep{isensee2021nnu} developed a nnU-Net, which incorporates a preset DA pipeline consisting of varying operations, including rotation, scaling, Gaussian noise, Gaussian blur, etc., in sequence. These approaches heavily rely on pipeline design experience and may lead to suboptimal augmentation diversity.

A notable proportion of researchers employ synthetic models such as GAN to synthesize artificial images. For instance, Beers et al. \citep{beers2018high} leveraged PGGAN \citep{karras2017progressive} to synthesize fundus and glioma images. Chaitanya et al. \citep{chaitanya2021semi} introduced a task-driven DA approach termed STDA, in which the synthetic generator models intensity and shape through additive intensity transformations and deformation fields. Chai et al. \citep{chai2022synthetic} developed a DPGAN to synthesize images and labels for vestibular schwannoma, kidney tumors, and skin cancer. The DPGAN comprises three variational auto-encoder GANs and an extra discriminator to enhance image reality and correlation among images and latent vectors. GAN-based approaches are computationally expensive, require large amounts of data, and may output varying-quality synthesis. Several studies leverage the diffusion model as an alternative. For example, Moghadam et al. \citep{moghadam2023morphology} generated histopathology images by employing diffusion models with color normalization and prioritized morphology weighting. Pinaya et al. \citep{pinaya2022brain} leveraged latent diffusion models \citep{rombach2022high} to synthesize three-dimensional artificial brain images. Tang et al. \citep{tang2023multi} employed latent diffusion models to synthesize unlabeled data for semi-supervised segmentation. Such methods can be constrained by low sampling speed and high computational cost.

Several researchers have leveraged the automatic DA approach. For instance, Qin et al. \citep{qin2020automatic} developed a joint-learning strategy to combine segmentation modules and Dueling DQN \citep{wang2016dueling} to search for maximum performance improvement. Xu et al. \citep{xu2020automatic} proposed a differentiable way to update the parameters using stochastic relaxation and the Monte Carlo method. Lyu et al. \citep{lyu2022aadg} introduced an AADG framework consisting of a new proxy task to maximize the diversity among various augmented domains using Sinkhorn distance. Additionally, Yang et al. \citep{yang2019searching} utilized the validation accuracy to update the recurrent neural network controller. These approaches face high computation costs. To this end, we introduce a suitable yet general automatic DA approach termed MedAugment with negligible computational overhead.

%% ---------------------------------------- Methods ---------------------------------------- %%

\section{Methods}
\label{3}

\subsection{MedAugment}
\label{3.1}

\begin{figure}
    \centering
    \includegraphics[width=0.6\linewidth]{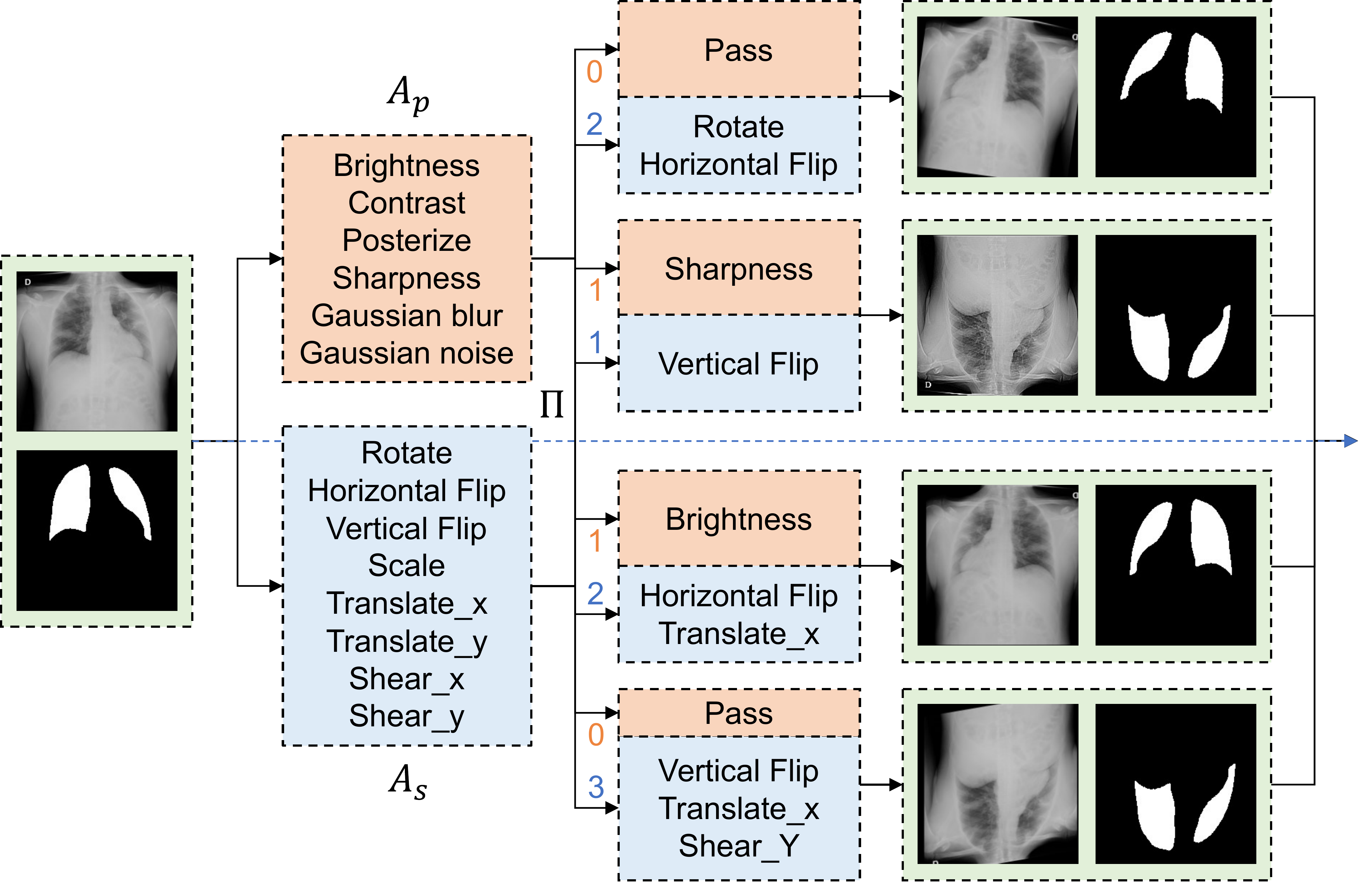}
    \caption{A realization of MedAugment. It comprises $N = 4$ augmentation branches and a separate branch to retain original input features. Each input generates five output images, comprising four augmented and one unaltered original. For each augmentation branch, $M = {\{2, 3}\}$ DA operations are sampled using the sampling strategy $\Pi$ from the pixel augmentation space $A_{p}$ and spatial augmentation space $A_{s}$.}
    \label{fig1}
\end{figure}

We illustrate a realization of the proposed MedAugment in \Cref{fig1}. The MedAugment encompasses $N = {4}$ augment branches and a separate branch to retain the input features. We design pixel augmentation space $A_{p}$ and spatial augmentation space $A_{s}$ and exclude the operations that can disrupt medical details and features. This results in six and eight DA operations in $A_{p}$ and $A_{s}$, respectively. Besides, we develop an operation sampling strategy $\Pi$ to restrict the number of operations sampled from the two spaces, resulting in $M = {\{2, 3}\}$ sequential DA operations in each augment branch. Moreover, we propose a mapping relationship to produce a rational augmentation level and affirm that the maximum magnitude $M_{A}$ and probability $P_{A}$ for each operation are controllable with a single augmentation level $l$. These designs can effectively handle the differences between natural and medical images. It is worth pointing out that several operations, such as \texttt{horizontal flip}, do not possess magnitude. 

\begin{algorithm}[!pt]
\small
\begin{algorithmic}[1]
\caption{Pseudocode for MedAugment.}
\label{algorithm1}  
\Require Pixel augmentation space $A_{p}$ = \texttt{\{brightness,..., gaussian noise}\}, spatial augmentation space $A_{s}$ = \texttt{\{rotate,..., shear\_y}\}, augmentation branch $B = {\{b_1,..., b_4\}}$, number of sequential operations $M = {\{2, 3}\}$, sampling strategy $\Pi = {\{\pi_1,..., \pi_4\}}$, augmentation level $l=5$, maximum operation magnitude $M_{A_{p}} = {\{0.1l,..., -}\}$, $M_{A_{s}} = {\{4l,..., (0, 0.02l)}\}$, operation probability $P_{A} = 0.2l$, input dataset $D = (X, Y)$;
\Ensure Augmented dataset $D^{a}$, output dataset $D^{o}$;
\ForAll {$X_i, Y_i$}
    \ForAll {$b_j$}
        \State Sample $\pi$ from $\Pi$ without replacement \Comment{\texttt{strategy-level random}}
        \State Sample $M$ operations $\mathcal{O}_j = {\{o_1, ..., o_{M}}\}$ using $\pi$ from $A$;
        \State Shuffle $\mathcal{O}_j$ \Comment{\texttt{operation-level random}}
            \ForAll {$o$}
                \State Calculate $M_{A}$, $P_{A}$ using $l$
                \State Sample magnitude \(m_A \sim \mathrm{UNIFORM}(M_A)\) \Comment{\texttt{magnitude-level random}}
            \EndFor
            \State $(X^j_i, Y^j_i) = \mathcal{O}_j(X_i, Y_i)$
            \State Add $(X^j_i, Y^j_i)$ to $D^{a}$
    \EndFor
\EndFor
\State Out $D^{o} = D^{a} + D$
\end{algorithmic}
\end{algorithm}

\begin{figure}
    \centering
    \includegraphics[width=0.6\linewidth]{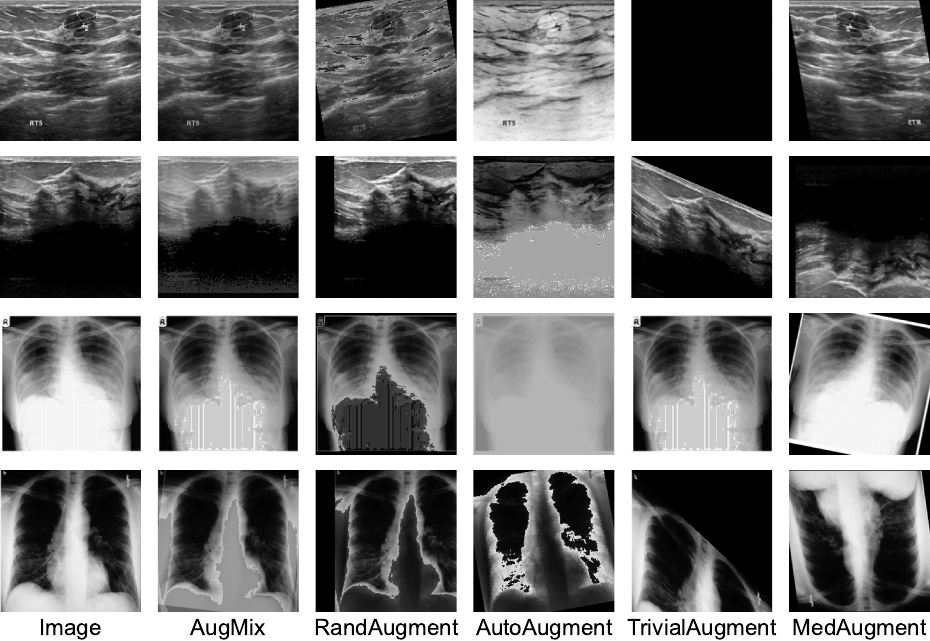}
    \caption{Examples of augmented images generated by varying automatic DA methods.}
    \label{fig2}
\end{figure}

We illustrate the pseudo-code of MedAugment in \Cref{algorithm1}. As demonstrated, MedAugment introduces randomness from three aspects, including the strategy level, operation level, and magnitude level. For each input image, MedAugment samples the operations for each branch using the sampling strategy. The sampled operations are then shuffled. Afterward, the input image is augmented with the shuffled operations, in which the operation magnitude is uniformly sampled within the maximum magnitude based on $l$. We present several examples of augmented images generated by different automatic DA methods in \Cref{fig2}. It can be observed that MedAugment can prevent severe color distortions or structural alterations and produce the most realistic augmented images. The augmented images generated by the remaining approaches may be identifiable to the DL model but can lack clinical relevance or interpretability.

\subsection{Augmentation Spaces}
\label{3.2}

We partition the DA operations into pixel-level and spatial-level operations, where pixel-level operations modify pixel values without changing their image structure, and spatial-level operations involve geometric transformations that alter the spatial arrangement of pixels. Based on this distinction, we construct $A_{p}$ and $A_{s}$ with pixel-level and spatial-level DA operations, respectively. To secure the eligibility of the proposed MedAugment for medical images, we exclude operations that can disrupt the details and features in medical images, such as \texttt{invert}, \texttt{equalize}, and \texttt{solarize}, based on pre-stage experiments. This results in the $A_{p}=$ \texttt{\{brightness, contrast, posterize, sharpness, gaussian blur, gaussian noise}\}, and $A_{s}=$ \texttt{\{rotate, horizontal flip, vertical flip, scale, translate\_x, translate\_y, shear\_x, shear\_y}\}. To prevent the operations in $A_{p}$ from hampering the grey-level class information in the masks, we solely employ the operations from $A_{s}$ for mask augmentation. To affirm compatibility and scalability, we leverage the well-established augmentation framework Albumentations \citep{buslaev2020albumentations} to perform conventional operations for its superior diversity \citep{perez2021torchio, fabian2020batchgenerators}.

\subsection{Sampling Strategy}
\label{3.3}

We present a $\Pi$ when sampling operations from $A_{p}$ and $A_{s}$ due to considerations from two aspects. Firstly, the medical images are sensitive to attributes such as \texttt{brightness}. Secondly, numerous consecutive operations can lead to unrealistic output images that drift far from the original ones \citep{hendrycks2019augmix}. To this end, we regulate the maximum number of operations sampled from $A_{p}$ and in total equals one and three, respectively. Besides, setting the total number of operations to one is considered inconsequential as it degrades to a single operation without combinations. Considering these factors, we affirm the number of sequential operations $M={\{2, 3}\}$. Given this setup, four sampling combinations $\Pi = {\{\pi_1, \pi_2, \pi_3, \pi_4\}}$ are produced, in which the number of the operations sampled from the two spaces equal $1+2$, $0+3$, $1+1$, and $0+2$, respectively. This number of combinations determines the number of augment branches. To ensure the scalability of MedAugment, we design the $N$ extendable with the sampling altered as replacement sampling. Besides, the separate branch can be shielded. By setting $N=1$ and shielding the separate branch, the MedAugment can be leveraged to perform one-to-one augmentation. 

\subsection{Hyperparameter Mapping}
\label{3.4}

We propose a hyperparameter mapping relationship to produce a rational augmentation level and make the $M_{A}$ and $P_{A}$ of each operation fully controllable using $l=\{1, 2, 3, 4, 5\}$, in which higher values correspond to stronger augmentation. We observed that medical images are susceptible to the magnitude of several operations like \texttt{posterize}. When the number of remaining bits decreases, the quality of the augmented images deteriorates substantially. Therefore, we meticulously design the magnitude of these types of operations based on extensive pre-stage experiments to affirm that the resultant augmented images retain their significance. We illustrate the mapping between $l$ and $M_{A}$ for different operations in \Cref{tab1}. Note that operations without magnitude are indicated as $-$. The function $F$ returns an odd number based on the input $l$ and is formulated as:
\begin{equation}
F(x)=\left\{\begin{array}{ll}\lceil x\rceil+1 & \lceil x\rceil=2 k \\ \lceil x\rceil & \lceil x\rceil=2 k+1\end{array} \quad k \in Z\right.
\label{eq1}
\end{equation}
where $\lceil \rceil$ represents round up. The probability for operations sampled from $A_{p}$ and $A_{s}$ adhere to the identical formulation, in which $P_{A} = 0.2l$.

\begin{table}[ht]
\centering
\caption{$M_{A}$ for operations from $A_{p}$ and $A_{s}$. $\lfloor \rfloor$ represents round down. The function $F$ returns an odd integer. $-$ denotes the operation without magnitude.}
\resizebox{0.6\linewidth}{!}{
\begin{tabular*}{416pt}{cccc}
\toprule
    Space                    & Operation       & Magnitude                   & Parameter \\
\midrule
    \multirow{6}{*}{$A_{p}$} & Brightness      & $0.04l$                     & Brightness \\
                             & Contrast        & $0.04l$                     & Contrast \\
                             & Posterize       & $\lfloor 8 - 0.8l \rfloor$  & Number of bits left \\
                             & Sharpness       & $(0.04l, 0.1l)$             & Sharpened image visibility \\
                             & Gaussian blur   & $(3, F(3+0.8l))$            & Maximum Gaussian kernel size \\
                             & Gaussian noise  & $(2l, 10l)$                 & Gaussian noise variance range \\
\midrule
    \multirow{8}{*}{$A_{s}$} & Rotate          & $4l$                        & Rotation in degree \\
                             & Horizontal flip & $-$                         & Horizontal flip \\
                             & Vertical flip   & $-$                         & Vertical flip \\
                             & Scale           & $(1-0.04l, 1+0.04l)$        & Scaling factor \\
                             & Translate\_x    & $(0, 2l)$                   & X translate in fraction \\
                             & Translate\_y    & $(0, 2l)$                   & Y translate in fraction \\
                             & Shear\_x        & $(0, 0.02l)$                & X shear in degree \\
                             & Shear\_y        & $(0, 0.02l)$                & Y shear in degree \\
\bottomrule
\end{tabular*}
}
\label{tab1}
\end{table}

%% ------------------------------- Experiments ------------------------------- %%

\section{Experiments}
\label{4}

\subsection{Datasets}
\label{4.1}

We leverage four datasets for classification performance evaluation. The breast ultrasound (BUSI) dataset \cite{al2020dataset} was collected from 600 female patients between 25 and 75 years old. It encompasses 780 images, of which 437, 210, and 133 are benign, malignant, and normal, respectively. The average image resolution of BUSI is around 500 $\times$ 500. The ultrasound nasogastric tube (UNGT) dataset \citep{liu2025ungt} is a nasogastric tube placement confirmation dataset extended from SNGT \cite{liu2024segmenting}. It includes 493 images gathered from 110 patients with an average image resolution of approximately 879 $\times$ 583. The lung diseases X-ray (LUNG) dataset \cite{tahir2022deep} was collected by Qatar University and the University of Dhaka. It has COVID-19, severe acute respiratory syndrome, and Middle East Respiratory Syndrome categories, with 423, 134, and 144 images each. The brain tumor magnetic resonance imaging (BTMRI) dataset \cite{btmri} comprises categories including glioma, meningioma, normal, and pituitary. Each category encompasses 1321, 1339, 1595, and 1457 images for training and 300, 306, 405, and 300 for testing.

We utilize four datasets for segmentation performance comparison, including LUNG, in which images and masks across categories are merged. The COVID-19 computed tomography scan lesion segmentation (COVID) dataset \cite{covid} consists of 2729 images and ground truth (GT) mask pairs. These images were curated from three public computed tomography sources, such as MosMedData \cite{morozov2020mosmeddata}. The endoscopic colonoscopy (CVC) dataset \cite{cvcdb} serves as the official database of the MICCAI training stages and contains 1224 polyp frames and masks extracted from colonoscopy videos. The colonoscopy image (Kvasir) dataset \cite{kvasir} is composed of 1000 gastrointestinal polyp images and masks with a resolution varying from 332 $\times$ 487 to 1920 $\times$ 1072.

\subsection{Experimental Setup}
\label{4.2}

We pre-process the datasets to 224 $\times$ 224 resolution. We divide the datasets into training, validation, and test subsets with a ratio of 6:2:2 or 8:2 in case testing data is separately provided. For classification datasets, the proportion of each category in different subsets equals that of the total. The class-balanced partition can prevent potential category imbalance. We augment the training subset across all methods to five times the original, following the one-to-five augmentation strategy of MedAugment. We term the approach without performing DA as NoAugment. We report the mean and standard deviation results across three independent runs.

For classification, we leverage the Adam optimizer with a 0.01 decay factor. We use cross-entropy loss with an initial learning rate of 0.002. The learning rate decays step-wise for every 20 epochs with a factor of 0.9. The total epoch is 40, and the early stopping technique is introduced with a patience of 8. The batch size is 128. We use convolution-based ResNet \cite{he2016deep} and attention-based SwT \cite{liu2021swin} for training. Models are evaluated based on metrics including accuracy (ACC), negative predictive value (NPV), positive predictive value (PPV), sensitivity (SEN), specificity (SPE), and F1 score (FOS). We compare our MedAugment with the state-of-the-art (SOTA) GAN-based GSDA \citep{liu2023gsda} and automatic DA methods including AugMix, RandAugment, AutoAugment, and TrivialAugment. Results reported are the mean across all categories when applicable.

Regarding segmentation, we utilize SoftIoU loss, and the remaining follow the classification setup. We leverage convolution-based UNet \cite{ronneberger2015u} and attention-based SwinUNETR \cite{hatamizadeh2021swin} for training \cite{iakubovskii2019segmentation}. We evaluate the model performance using dice score (DSC), intersection over union (IoU), and pixel accuracy (PAC). We compare the MedAugment with the SOTA GAN-based STDA and varying automatic DA methods. As most automatic DA methods \citep{lingchen2020uniformaugment, hendrycks2019augmix, yun2019cutmix} are initially designed for classification, introducing them to segmentation can face mask misalignment by image mixture or requires method modifications. To this end, we propose unique conventional pipelines as SOTA approaches for performance comparison. Following the report \citep{eisenmann2023winner} that horizontal flip, rotate, and vertical flip are the most prevalent implemented operations in MIA, we present MonoAugment, DuoAugment, and TriAugment. The MonoAugment encompasses solely \texttt{horizontal flip}, while the DuoAugment and TriAugment are composed of successive \texttt{horizontal flip, rotate} and \texttt{horizontal flip, rotate, vertical flip}, respectively. The probability for each DA operation $p=0.5$.

%% ------------------------------- Results and Analysis ------------------------------- %%

\section{Results and Analysis}
\label{5}

\subsection{Classification}
\label{5.1}

\begin{table}[ht!]
\centering
\caption{Classification results across datasets using ResNet. NoAugment stands for results without DA. The best results are in bold. Superscripts on MedAugment denote p-values from the Wilcoxon signed-rank test comparing MedAugment with the best-performing baseline across three runs. Note that with three runs, the minimum achievable p-value is 0.25, and statistical significance at the 0.05 level cannot be reached.}
\resizebox{\linewidth}{!}{
\begin{tabular*}{674pt}{ccccccccc}
\toprule
    Dataset                & Metrics  & NoAugment               & GSDA                    & AugMix                  & RandAugment             & AutoAugment             & TrivialAugment          & MedAugment              \\
\midrule
    \multirow{6}{*}{BUSI}  
                           & ACC      & 58.37$\pm$7.11          & 61.57$\pm$3.06          & 70.67$\pm$2.37          & 62.20$\pm$3.48          & 70.73$\pm$2.37          & 69.00$\pm$3.48          & \textbf{76.00}$^{0.25}\pm$1.65 \\
                           & NPV      & 75.97$\pm$11.56         & 79.20$\pm$4.92          & 84.47$\pm$1.60          & 78.63$\pm$4.70          & 85.67$\pm$1.77          & 82.83$\pm$1.83          & \textbf{86.73}$^{0.50}\pm$1.07 \\
                           & PPV      & 58.50$\pm$14.38         & 62.77$\pm$7.76          & 70.63$\pm$1.93          & 55.60$\pm$11.52         & 65.73$\pm$11.13         & 66.33$\pm$4.00          & \textbf{75.43}$^{0.25}\pm$3.18 \\
                           & SEN      & 50.37$\pm$15.05         & 51.27$\pm$11.94         & 62.67$\pm$5.36          & 49.50$\pm$7.43          & 61.80$\pm$8.50          & 66.70$\pm$6.07          & \textbf{70.63}$^{0.75}\pm$2.01 \\
                           & SPE      & 74.13$\pm$7.40          & 75.17$\pm$4.58          & 81.27$\pm$2.21          & 74.10$\pm$3.83          & 81.80$\pm$3.69          & 82.27$\pm$2.01          & \textbf{86.03}$^{0.25}\pm$0.52 \\
                           & FOS      & 46.73$\pm$12.57         & 47.97$\pm$11.04         & 63.47$\pm$3.84          & 45.83$\pm$6.48          & 59.07$\pm$7.97          & 64.33$\pm$4.07          & \textbf{71.30}$^{0.50}\pm$2.00 \\
\midrule
    \multirow{6}{*}{UNGT}  
                           & ACC      & 73.33$\pm$1.89          & 75.00$\pm$5.10          & 81.33$\pm$3.40          & 82.33$\pm$0.94          & 82.00$\pm$0.82          & 82.67$\pm$2.62          & \textbf{84.00}$^{1.00}\pm$0.82 \\
                           & NPV      & 74.27$\pm$1.77          & 80.77$\pm$3.94          & 81.23$\pm$2.66          & 82.60$\pm$1.27          & 82.43$\pm$3.64          & 83.67$\pm$2.62          & \textbf{84.00}$^{0.75}\pm$2.12 \\
                           & PPV      & 74.27$\pm$1.77          & 80.77$\pm$3.94          & 81.23$\pm$2.66          & 82.60$\pm$1.27          & 82.43$\pm$3.64          & 83.67$\pm$2.62          & \textbf{84.00}$^{0.75}\pm$2.12 \\
                           & SEN      & 74.63$\pm$2.53          & 68.00$\pm$9.85          & 80.17$\pm$5.85          & 78.27$\pm$2.88          & 78.27$\pm$1.34          & 79.00$\pm$5.39          & \textbf{81.53}$^{1.00}\pm$2.19 \\
                           & SPE      & 74.63$\pm$2.53          & 68.00$\pm$9.85          & 80.17$\pm$5.85          & 78.27$\pm$2.88          & 78.27$\pm$1.34          & 79.00$\pm$5.39          & \textbf{81.53}$^{1.00}\pm$2.19 \\
                           & FOS      & 71.93$\pm$0.41          & 66.10$\pm$12.81         & 79.20$\pm$4.63          & 79.33$\pm$2.22          & 79.20$\pm$0.37          & 79.63$\pm$4.22          & \textbf{81.97}$^{0.75}\pm$1.07 \\
\midrule
    \multirow{6}{*}{LUNG}  
                           & ACC      & 63.33$\pm$0.66          & 72.10$\pm$1.48          & 66.93$\pm$0.87          & 72.13$\pm$7.19          & 71.63$\pm$4.36          & 73.73$\pm$1.04          & \textbf{77.53}$^{0.25}\pm$2.31 \\
                           & NPV      & 82.90$\pm$5.00          & 85.60$\pm$1.23          & 86.90$\pm$2.26          & 86.00$\pm$3.40          & 87.27$\pm$3.40          & 87.60$\pm$1.84          & \textbf{88.80}$^{0.75}\pm$0.86 \\
                           & PPV      & 48.10$\pm$6.14          & 56.40$\pm$12.46         & 52.97$\pm$2.45          & 64.17$\pm$16.20         & 75.57$\pm$14.95         & 79.80$\pm$6.30          & \textbf{80.33}$^{0.75}\pm$0.66 \\
                           & SEN      & 42.60$\pm$4.46          & 57.30$\pm$3.91          & 44.80$\pm$2.45          & 59.13$\pm$12.34         & 53.50$\pm$6.38          & 59.50$\pm$3.23          & \textbf{66.23}$^{0.25}\pm$5.03 \\
                           & SPE      & 71.73$\pm$1.72          & 79.40$\pm$1.57          & 72.50$\pm$1.16          & 81.13$\pm$6.21          & 77.60$\pm$2.82          & 80.20$\pm$1.96          & \textbf{84.17}$^{0.50}\pm$2.07 \\
                           & FOS      & 38.87$\pm$3.58          & 55.13$\pm$6.54          & 42.90$\pm$2.55          & 56.73$\pm$13.69         & 54.50$\pm$9.12          & 62.40$\pm$1.82          & \textbf{66.73}$^{0.50}\pm$7.25 \\
\midrule
    \multirow{6}{*}{BTMRI} 
                           & ACC      & 82.90$\pm$2.43          & 90.43$\pm$2.50          & 91.33$\pm$1.11          & 91.90$\pm$0.64          & 92.53$\pm$0.38          & 93.47$\pm$0.21          & \textbf{94.70}$^{0.25}\pm$0.29 \\
                           & NPV      & 94.63$\pm$0.81          & 96.87$\pm$0.84          & 97.23$\pm$0.33          & 97.37$\pm$0.19          & 97.53$\pm$0.17          & 97.83$\pm$0.05          & \textbf{98.30}$^{0.25}\pm$0.08 \\
                           & PPV      & 84.30$\pm$3.44          & 90.90$\pm$2.24          & 91.90$\pm$1.00          & 92.00$\pm$0.41          & 92.47$\pm$0.54          & 93.57$\pm$0.05          & \textbf{94.63}$^{0.25}\pm$0.21 \\
                           & SEN      & 81.90$\pm$2.49          & 90.33$\pm$2.32          & 90.77$\pm$1.09          & 91.43$\pm$0.80          & 92.13$\pm$0.25          & 93.30$\pm$0.42          & \textbf{94.37}$^{0.25}\pm$0.40 \\
                           & SPE      & 94.30$\pm$0.88          & 96.80$\pm$0.85          & 97.10$\pm$0.36          & 97.30$\pm$0.28          & 97.53$\pm$0.09          & 97.83$\pm$0.09          & \textbf{98.23}$^{0.25}\pm$0.09 \\
                           & FOS      & 81.67$\pm$2.78          & 90.33$\pm$2.41          & 91.00$\pm$1.13          & 91.57$\pm$0.62          & 92.17$\pm$0.34          & 93.30$\pm$0.22          & \textbf{94.40}$^{0.25}\pm$0.36 \\
\bottomrule
\end{tabular*}
}
\label{tab2}
\end{table}

\begin{table}[ht!]
\centering
\caption{Classification results across datasets using SwT.}
\resizebox{\linewidth}{!}{
\begin{tabular*}{674pt}{ccccccccc}
\toprule
    Dataset                & Metrics  & NoAugment               & GSDA                    & AugMix                  & AutoAugment             & RandAugment             & TrivialAugment          & MedAugment              \\
\midrule
    \multirow{6}{*}{BUSI}  
                           & ACC      & 81.07$\pm$2.10          & 81.53$\pm$0.53          & 82.37$\pm$1.46          & 82.80$\pm$1.37          & 83.00$\pm$1.31          & 83.23$\pm$0.78          & \textbf{84.10}$^{0.50}\pm$1.55 \\
                           & NPV      & 89.60$\pm$1.04          & 89.57$\pm$0.33          & 89.90$\pm$0.71          & 90.73$\pm$0.63          & 90.60$\pm$0.64          & \textbf{91.43}$\pm$0.34 & 91.00$\pm$0.70          \\
                           & PPV      & 78.57$\pm$1.92          & 79.57$\pm$0.60          & 79.57$\pm$0.99          & 82.27$\pm$1.86          & 82.17$\pm$1.13          & \textbf{83.87}$\pm$0.78 & 83.00$\pm$0.71          \\
                           & SEN      & 77.07$\pm$2.99          & 79.47$\pm$0.71          & 81.83$\pm$3.50          & 79.40$\pm$3.02          & 80.23$\pm$2.47          & 78.23$\pm$1.96          & \textbf{82.40}$^{0.50}\pm$2.79 \\
                           & SPE      & 88.97$\pm$1.55          & 89.17$\pm$0.33          & 90.43$\pm$1.60          & 89.50$\pm$1.61          & 89.57$\pm$1.09          & 89.10$\pm$0.78          & \textbf{90.60}$^{0.75}\pm$1.42 \\
                           & FOS      & 77.60$\pm$2.62          & 79.23$\pm$0.62          & 80.30$\pm$2.12          & 80.23$\pm$1.80          & 80.80$\pm$1.68          & 80.30$\pm$1.31          & \textbf{82.33}$^{0.25}\pm$1.72 \\
\midrule
    \multirow{6}{*}{UNGT}  
                           & ACC      & 87.33$\pm$2.36          & 90.00$\pm$0.00          & 89.67$\pm$1.25          & 88.67$\pm$0.47          & 89.33$\pm$2.05          & 87.67$\pm$1.89          & \textbf{90.67}$^{0.50}\pm$0.47 \\
                           & NPV      & 87.40$\pm$2.69          & 89.90$\pm$0.28          & 88.83$\pm$1.27          & 88.00$\pm$0.14          & 90.10$\pm$2.36          & 87.60$\pm$1.08          & \textbf{90.63}$^{1.00}\pm$0.38 \\
                           & PPV      & 87.40$\pm$2.69          & 89.90$\pm$0.28          & 88.83$\pm$1.27          & 88.00$\pm$0.14          & 90.10$\pm$2.36          & 87.60$\pm$1.08          & \textbf{90.63}$^{1.00}\pm$0.38 \\
                           & SEN      & 84.33$\pm$2.78          & 87.93$\pm$0.33          & 88.33$\pm$1.54          & 86.90$\pm$0.99          & 86.30$\pm$2.37          & 85.00$\pm$3.19          & \textbf{88.63}$^{0.75}\pm$0.66 \\
                           & SPE      & 84.33$\pm$2.78          & 87.93$\pm$0.33          & 88.33$\pm$1.54          & 86.90$\pm$0.99          & 86.30$\pm$2.37          & 85.00$\pm$3.19          & \textbf{88.63}$^{0.75}\pm$0.66 \\
                           & FOS      & 85.53$\pm$2.78          & 88.77$\pm$0.09          & 88.53$\pm$1.43          & 87.33$\pm$0.66          & 87.73$\pm$2.41          & 85.93$\pm$2.57          & \textbf{89.50}$^{0.50}\pm$0.57 \\
\midrule
    \multirow{6}{*}{LUNG}  
                           & ACC      & 84.40$\pm$0.00          & 86.97$\pm$0.66          & 86.97$\pm$0.66          & 86.73$\pm$0.33          & 84.87$\pm$0.33          & 86.03$\pm$0.66          & \textbf{88.20}$^{0.50}\pm$1.24 \\
                           & NPV      & 92.63$\pm$0.24          & 93.60$\pm$0.29          & 93.73$\pm$0.45          & 93.47$\pm$0.38          & 92.67$\pm$0.33          & 93.33$\pm$0.52          & \textbf{94.13}$^{0.25}\pm$0.59 \\
                           & PPV      & 89.57$\pm$0.61          & 89.57$\pm$0.40          & 90.17$\pm$0.97          & \textbf{90.47}$\pm$0.66 & 89.37$\pm$0.76          & 90.17$\pm$1.60          & 90.43$\pm$1.24          \\
                           & SEN      & 75.77$\pm$0.38          & 80.37$\pm$1.33          & 80.17$\pm$0.92          & 80.13$\pm$0.19          & 76.87$\pm$0.46          & 78.57$\pm$0.76          & \textbf{82.40}$^{0.50}\pm$1.88 \\
                           & SPE      & 87.60$\pm$0.14          & 90.20$\pm$0.65          & 90.00$\pm$0.62          & 89.70$\pm$0.14          & 88.10$\pm$0.24          & 89.10$\pm$0.28          & \textbf{91.10}$^{0.50}\pm$0.88 \\
                           & FOS      & 80.60$\pm$0.14          & 83.97$\pm$1.02          & 83.97$\pm$0.80          & 83.87$\pm$0.38          & 81.37$\pm$0.42          & 82.77$\pm$0.97          & \textbf{85.63}$^{0.50}\pm$1.75 \\
\midrule
    \multirow{6}{*}{BTMRI} 
                           & ACC      & 87.77$\pm$0.48          & 88.63$\pm$0.05          & 88.47$\pm$0.40          & 89.30$\pm$0.36          & 88.87$\pm$0.19          & 88.93$\pm$0.31          & \textbf{90.37}$^{0.25}\pm$0.24 \\
                           & NPV      & 96.00$\pm$0.14          & 96.30$\pm$0.00          & 96.23$\pm$0.09          & 96.50$\pm$0.14          & 96.37$\pm$0.05          & 96.40$\pm$0.08          & \textbf{96.83}$^{0.25}\pm$0.05 \\
                           & PPV      & 87.73$\pm$0.33          & 88.87$\pm$0.05          & 88.53$\pm$0.46          & 89.27$\pm$0.21          & 88.53$\pm$0.17          & 88.93$\pm$0.41          & \textbf{90.17}$^{0.25}\pm$0.17 \\
                           & SEN      & 87.23$\pm$0.52          & 88.13$\pm$0.05          & 87.90$\pm$0.42          & 88.80$\pm$0.42          & 88.47$\pm$0.19          & 88.43$\pm$0.39          & \textbf{89.93}$^{0.25}\pm$0.33 \\
                           & SPE      & 95.93$\pm$0.17          & 96.20$\pm$0.00          & 96.10$\pm$0.14          & 96.40$\pm$0.14          & 96.30$\pm$0.08          & 96.30$\pm$0.14          & \textbf{96.77}$^{0.25}\pm$0.09 \\
                           & FOS      & 87.27$\pm$0.48          & 88.33$\pm$0.05          & 88.07$\pm$0.40          & 88.93$\pm$0.38          & 88.43$\pm$0.17          & 88.53$\pm$0.39          & \textbf{89.97}$^{0.25}\pm$0.24 \\
\bottomrule
\end{tabular*}
}
\label{tab3}
\end{table}

We demonstrate the classification results in \Cref{tab2} and \Cref{tab3} for ResNet and SwT. As can be observed, the proposed MedAugment overperforms SOTA methods across the models. For ResNet, MedAugment ranks first in 24 metrics. On BUSI, MedAugment achieves the highest metrics with an ACC of 76.00\%. The GSDA does not present an ideal performance with a 61.57\% ACC. On UNGT, MedAugment realizes the optimal six metrics, achieving an ACC of 84.00\%. Similar observations can be found where GSDA falls behind the remaining approaches. On LUNG, MedAugment reaches the highest performance with a 77.53\% ACC. Additionally, the AugMix merely realizes an ACC of 66.93\%. On BTMRI, MedAugment achieves the optimal metrics with 94.70\% ACC. Consistently, GSDA presents the lowest ACC of 90.43\%. Regarding SwT, MedAugment ranks first in 21 out of 24 metrics. On BUSI, it achieves optimal ACC, SEN, SPE, and FOS of 84.10\%, 82.40\%, 90.60\%, and 82.33\%. Following MedAugment, TrivialAugment realizes the highest NPV and PPV of 91.43\% and 83.87\%. The lowest performance is observed for GSDA with an 81.53\% ACC. On UNGT, MedAugment realizes the highest six metrics with a 90.67\% ACC. Conversely, TrivialAugment does not present superior results with an ACC of 87.67\%. On LUNG, MedAugment achieves the highest ACC, NPV, SEN, SPE, and FOS of 88.20\%, 94.13\%, 82.40\%, 91.10\%, and 85.63\%, followed by 90.47\% PPV realized by AutoAugment. Additionally, RandAugment achieves a suboptimal performance with 84.87\% ACC. On BTMRI, MedAugment realizes the best performance across metrics with 90.37\% ACC. The GSDA, AugMix, RandAugment, and TrivialAugment present similar results under this configuration. It is worth noting that relatively low SEN can be observed on the LUNG dataset, suggesting the difficulties in identifying subtle differences across disease categories, especially at their early stage.

\begin{figure}
    \centering
    \includegraphics[width=0.6\linewidth]{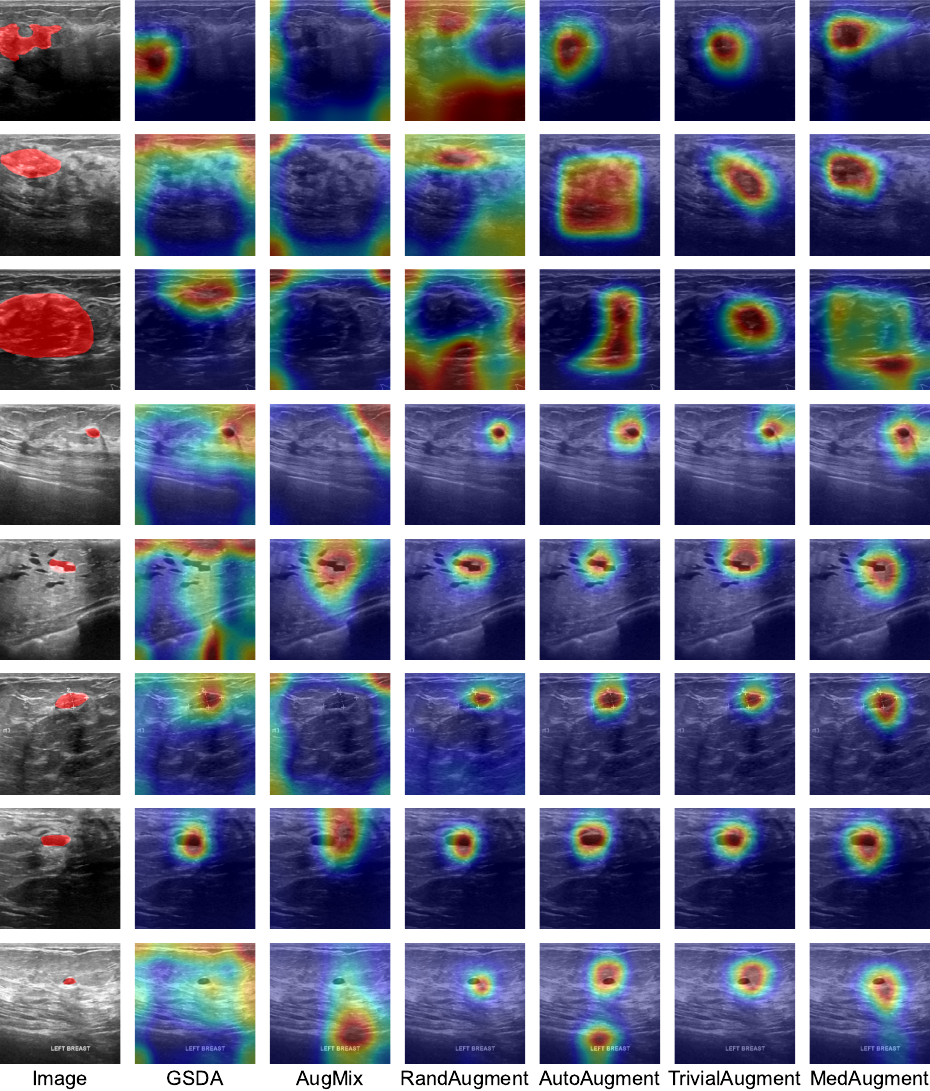}
    \caption{Class activation map across different DA methods on the BUSI dataset using ResNet. The red regions present the region of interest.}       
    \label{fig3}
\end{figure}

We employ the class activation map to evaluate the classification interpretability across different DA methods using ResNet in \Cref{fig3}. We leverage the BUSI dataset as its tumor regions are markedly discernible against the background. This enables an intuitive differentiation of the overlay between the class activation map and the image. We extract the features from layer 4, the final convolutional block of ResNet that encodes high-level semantic information, and employ the GradCAM method. Through observation, it is evident that the MedAugment can accurately focus on the correct regions, achieving fuller coverage and fitter contouring. In the first image, MedAugment allocates a substantial proportion of attention to the tumor region, whereas AugMix and RandAugment exhibit more dispersed attention. Similar patterns are observed in the second image, where decentralized attention is noted for GSDA, AugMix, and RandAugment. In addition, AutoAugment and TrivialAugment predominantly focus on unrelated regions. The third image is particularly challenging, as most methods demonstrate incorrect attention. Despite this, MedAugment performs better than the remaining methods, though the primary attention area shifts downward. In the sixth one, most methods successfully identify the tumor region, except for the AugMix, which focuses on the image boundary. Comparable results are observed in the seventh image, where AugMix underperforms the remaining ones. Regarding the last image, MedAugment demonstrates the most precise attention, followed by TrivialAugment and RandAugment. The other approaches either misallocate attention to irrelevant regions or distribute attention across multiple areas.

\begin{figure}
    \centering
    \includegraphics[width=0.6\linewidth]{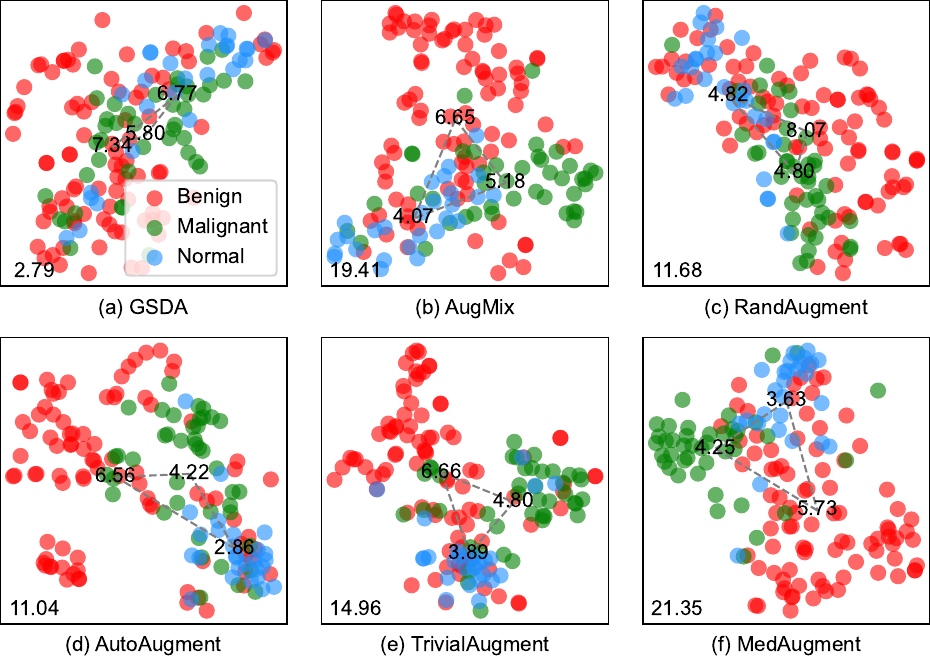}
    \caption{T-SNE visualization across different DA methods on the BUSI dataset using ResNet. The triangles are formed by connecting the centroids of each category. The numbers on the centroids indicate the standard deviations of points within each category, while the bottom-left number represents the area of the triangle across the three categories.}
    \label{fig4}
\end{figure}

We leverage t-SNE to visualize feature distributions and classification confidence across varying DA methods on the BUSI dataset using ResNet in \Cref{fig4}. We extract the features from layer 4. It can be observed that MedAugment achieves the most intensive intra-category distribution and dispersed inter-category distribution. For intra-category distribution, MedAugment achieves one of the most compact point distributions, with standard deviations of 5.73, 4.25, and 3.63 for the benign, malignant, and normal categories. This results in a total standard deviation of 13.61. Afterward, AutoAugment closely follows MedAugment with deviations equal to 6.56, 4.22, and 2.86, resulting in a total standard deviation of 13.64. In contrast, GSDA exhibits the least favorable results, with a total standard deviation equal to 19.91. Concerning inter-category distribution, MedAugment achieves the highest triangle area of 21.25 across the three categories. Subsequently, AugMix and TrivialAugment realize triangle areas of 19.41 and 14.96, respectively. However, GSDA demonstrates the least favorable performance, with a triangle area of only 2.79.

\subsection{Segmentation}
\label{5.2}

\begin{table}[ht!]
\centering
\caption{Segmentation results across datasets using UNet and SwinUNETR.}
\resizebox{\linewidth}{!}{
\begin{tabular*}{658pt}{ccccccccc}
\toprule
    Model 
    & Dataset                   & Metrics  & NoAugment               & STDA                    & MonoAugment             & DuoAugment              & TriAugment              & MedAugment              \\
\midrule
    \multirow{12}{*}{\raisebox{-1.5\height}{UNet}}
    & \multirow{3}{*}{LUNG}     & DSC      & 74.53$\pm$3.83          & 91.73$\pm$0.52          & 92.23$\pm$0.09          & 88.37$\pm$2.57          & 86.20$\pm$0.65          & \textbf{94.13}$^{0.25}\pm$0.31 \\
    &                           & IoU      & 62.03$\pm$4.15          & 85.30$\pm$0.86          & 86.07$\pm$0.26          & 80.23$\pm$3.89          & 76.93$\pm$0.97          & \textbf{89.27}$^{0.25}\pm$0.48 \\
    &                           & PAC      & 88.17$\pm$0.42          & 95.83$\pm$0.24          & 96.07$\pm$0.09          & 94.23$\pm$1.11          & 93.30$\pm$0.36          & \textbf{96.97}$^{0.25}\pm$0.12 \\
    \cmidrule(l{0pt}r{-1pt}){2-9}
    & \multirow{3}{*}{COVID}    & DSC      & 52.60$\pm$2.69          & 62.23$\pm$1.73          & 62.80$\pm$1.98          & 57.27$\pm$0.95          & 54.03$\pm$1.60          & \textbf{64.20}$^{0.50}\pm$0.49 \\
    &                           & IoU      & 41.00$\pm$2.28          & 49.90$\pm$1.71          & 50.53$\pm$1.89          & 44.93$\pm$0.53          & 41.53$\pm$1.89          & \textbf{51.97}$^{0.50}\pm$0.09 \\
    &                           & PAC      & 99.07$\pm$0.05          & 99.17$\pm$0.05          & 99.23$\pm$0.05          & 99.07$\pm$0.05          & 98.97$\pm$0.21          & \textbf{99.27}$^{1.00}\pm$0.05 \\
    \cmidrule(l{0pt}r{-1pt}){2-9}
    & \multirow{3}{*}{CVC}      & DSC      & 19.80$\pm$13.18         & 60.97$\pm$4.29          & 35.60$\pm$2.84          & 33.50$\pm$4.75          & 63.13$\pm$4.74          & \textbf{69.87}$^{0.25}\pm$3.78 \\
    &                           & IoU      & 13.77$\pm$9.13          & 50.17$\pm$3.39          & 27.37$\pm$2.52          & 22.77$\pm$4.23          & 51.67$\pm$4.69          & \textbf{59.37}$^{0.25}\pm$3.43 \\
    &                           & PAC      & 83.00$\pm$7.29          & 93.10$\pm$0.80          & 90.63$\pm$0.26          & 60.83$\pm$17.75         & 92.60$\pm$0.57          & \textbf{94.03}$^{0.50}\pm$0.87 \\
    \cmidrule(l{0pt}r{-1pt}){2-9}
    & \multirow{3}{*}{Kvasir}   & DSC      & 43.77$\pm$5.19          & 71.50$\pm$2.11          & 72.27$\pm$2.86          & 73.70$\pm$2.24          & 73.10$\pm$0.92          & \textbf{73.80}$^{1.00}\pm$0.24 \\
    &                           & IoU      & 30.80$\pm$4.70          & 60.53$\pm$1.75          & 61.27$\pm$3.19          & 62.50$\pm$2.55          & 61.77$\pm$1.31          & \textbf{63.03}$^{1.00}\pm$0.33 \\
    &                           & PAC      & 67.37$\pm$13.20         & 91.00$\pm$0.64          & 89.13$\pm$2.62          & 91.47$\pm$0.29          & \textbf{91.53}$\pm$0.45 & 91.00$\pm$0.62          \\
\midrule
    \multirow{12}{*}{\raisebox{-1.5\height}{SwinUNETR}} 
    & \multirow{3}{*}{LUNG}     & DSC      & 91.97$\pm$0.17          & 93.87$\pm$0.12          & 93.00$\pm$0.08          & 92.80$\pm$0.08          & 91.67$\pm$0.26          & \textbf{94.10}$^{0.50}\pm$0.22 \\
    &                           & IoU      & 85.67$\pm$0.25          & 88.77$\pm$0.12          & 87.43$\pm$0.12          & 87.03$\pm$0.12          & 85.20$\pm$0.36          & \textbf{89.20}$^{0.25}\pm$0.43 \\
    &                           & PAC      & 95.77$\pm$0.09          & 96.73$\pm$0.05          & 96.37$\pm$0.05          & 96.23$\pm$0.05          & 95.70$\pm$0.14          & \textbf{96.87}$^{0.50}\pm$0.09 \\
    \cmidrule(l{0pt}r{-1pt}){2-9}
    & \multirow{3}{*}{COVID}    & DSC      & 66.10$\pm$0.41          & 67.33$\pm$0.05          & 68.27$\pm$0.39          & 67.73$\pm$0.34          & 65.57$\pm$0.17          & \textbf{68.37}$^{0.50}\pm$0.31 \\
    &                           & IoU      & 53.27$\pm$0.56          & 54.57$\pm$0.09          & \textbf{55.40}$\pm$0.49 & 54.83$\pm$0.34          & 52.40$\pm$0.29          & 55.37$\pm$0.37          \\
    &                           & PAC      & 99.17$\pm$0.05          & 99.13$\pm$0.05          & \textbf{99.20}$\pm$0.00 & 99.10$\pm$0.00          & 99.10$\pm$0.00          & \textbf{99.20}$^{1.00}\pm$0.00 \\
    \cmidrule(l{0pt}r{-1pt}){2-9}
    & \multirow{3}{*}{CVC}      & DSC      & 61.90$\pm$1.34          & 72.30$\pm$0.49          & 72.43$\pm$0.33          & 72.50$\pm$0.28          & 60.17$\pm$0.34          & \textbf{76.27}$^{0.25}\pm$1.02 \\
    &                           & IoU      & 48.87$\pm$1.53          & 60.60$\pm$0.80          & 61.17$\pm$0.45          & 61.00$\pm$0.71          & 48.27$\pm$0.54          & \textbf{65.67}$^{0.25}\pm$1.48 \\
    &                           & PAC      & 91.00$\pm$0.43          & 93.60$\pm$0.14          & 93.70$\pm$0.08          & 93.73$\pm$0.05          & 91.30$\pm$0.43          & \textbf{94.53}$^{0.25}\pm$0.21 \\
    \cmidrule(l{0pt}r{-1pt}){2-9}
    & \multirow{3}{*}{Kvasir}   & DSC      & 50.93$\pm$0.39          & 57.37$\pm$0.45          & 57.97$\pm$0.66          & 51.67$\pm$0.45          & 54.37$\pm$0.40          & \textbf{60.67}$^{0.25}\pm$0.69 \\
    &                           & IoU      & 37.07$\pm$0.29          & 44.03$\pm$0.40          & 44.43$\pm$0.66          & 37.77$\pm$0.37          & 40.53$\pm$0.46          & \textbf{47.50}$^{0.25}\pm$0.78 \\
    &                           & PAC      & 80.27$\pm$0.95          & 85.80$\pm$0.45          & 85.60$\pm$0.29          & 81.30$\pm$0.94          & 83.67$\pm$0.79          & \textbf{87.03}$^{0.25}\pm$0.40 \\
\bottomrule
\end{tabular*}
}
\label{tab4}
\end{table}

We show the segmentation results for UNet and SwinUNETR in \Cref{tab4}. It is observable that MedAugment achieves the best performance compared with the remaining approaches. For UNet, MedAugment ranks first in 11 of 12 metrics. On LUNG, it achieves the highest metrics with a DSC of 94.13\%. Relatively poor performance is observed for TriAugment with a DSC of 86.20\%. On COVID, MedAugment realizes the optimal metrics with a DSC of 64.20\%. Consistently, the TriAugment does not demonstrate a superior performance with a 54.03\% DSC. On CVC, MedAugment achieves the best metrics with a 69.87\% DSC. Notably, the MonoAugment and DuoAugment methods exhibit significantly lower performance with a DSC equal to 35.60\% and 33.50\%. On Kvasir, MedAugment achieves the highest DSC and IoU of 73.80\% and 63.03\%. Afterward, TriAugment realizes the optimal PAC of 91.53\%. The STDA does not perform well with a DSC of 71.50\%. Concerning SwinUNETR, MedAugment ranks first in 11 of 12 metrics. On LUNG, MedAugment achieves the best metrics with a DSC of 94.10\%. TriAugment does not present an ideal performance with a DSC of 91.67\%, slightly lower than the results without DA. On COVID, MedAugment reaches the highest DSC and PAC of 68.37\% and 99.20\%. Subsequently, MonoAugment realizes the best IoU and PAC of 55.40\% and 99.20\%. TriAugment presents the lowest performance under this setup. On CVC, MedAugment achieves the optimal results with a 76.27\% DSC. Consistently, the TriAugment slightly underperforms the model trained without DA. On Kvasir, MedAugment achieves the best metrics with a 60.67\% DSC, and DuoAugment presents a relatively poor DSC of 51.67\%. Notably, significantly high PAC can be observed compared with DSC and IoU. However, such an observation may not indicate superior model performance as the predicted pixels may be dispersed. Moreover, the model may attain an ideal PAC even if the object is mistakenly predicted as the background when the object size is much smaller than the background.

\begin{figure}
    \centering
    \includegraphics[width=0.6\linewidth]{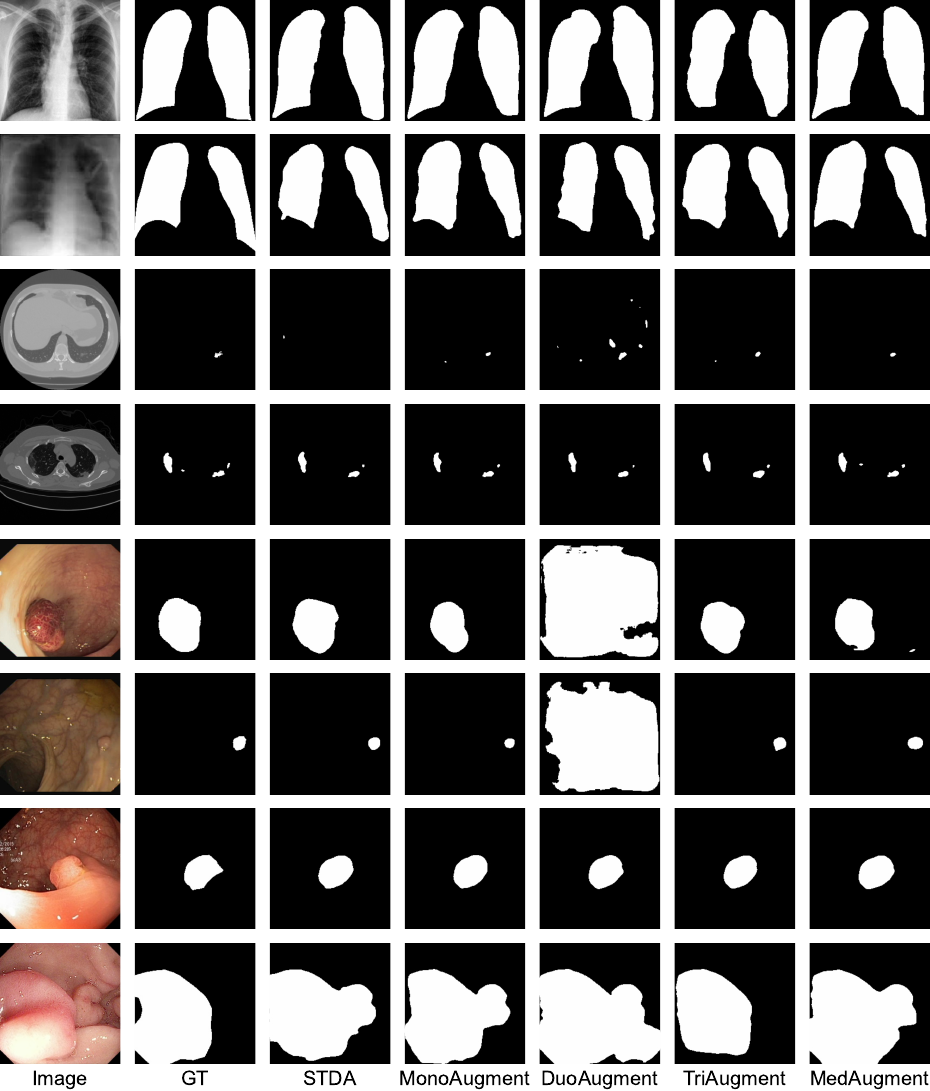}
    \caption{Predicted and GT masks across different DA methods on varying datasets using UNet.}
    \label{fig5}
\end{figure}

We present the predicted and GT masks across varying methods for different datasets using UNet in \Cref{fig5}. The comparison demonstrates that the MedAugment achieves the fewest erroneously predicted pixels and the highest contour prediction accuracy. In the first image, MedAugment achieves one of the most accurate lung contour predictions, especially for the left lung. In contrast, TriAugment fails to predict accurate contours. A similar observation is made in the second image, where MedAugment exhibits one of the optimal methods and TriAugment shows relatively poor results. In the third image, MedAugment largely outperforms the other methods, as they either predict incorrect regions or generate excessive areas, particularly for DuoAugment. For the sixth image, most methods yield accurate predictions except for the DuoAugment which largely overestimates the region. In the next one, all methods produce satisfactory predictions with minor performance differences. In the last image, prediction accuracy across methods is acceptable but not outstanding. Specifically, STDA, MonoAugment, DuoAugment, and MedAugment overestimate the region, while TriAugment underestimates it.

\begin{figure*}
    \includegraphics[width=\textwidth]{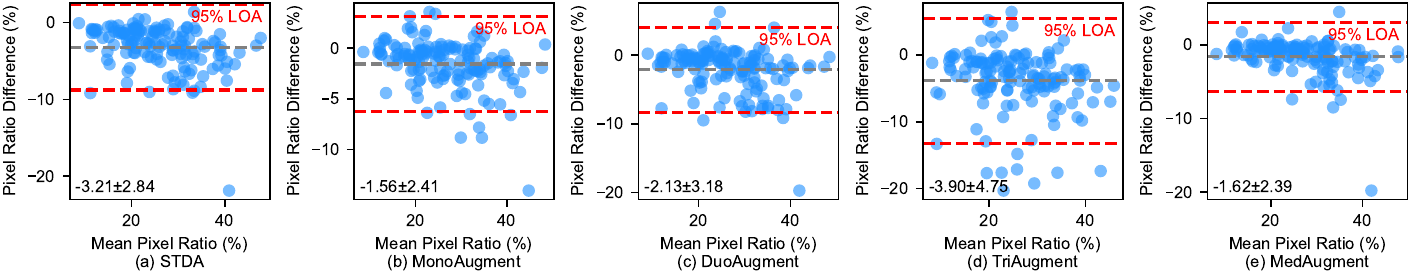}
    \caption{Bland-Altman plot across different DA methods on the LUNG dataset using UNet. LOA denotes the limits of agreement.}
    \label{fig6}
\end{figure*}

We leverage the Bland-Altman plot to illustrate the relationship between prediction-GT pairs across different DA methods on the LUNG dataset using ResNet in \Cref{fig6}. We count the number of object pixels in the predicted masks, divide the results by the total pixels, and compare the output with the corresponding GT masks. The results indicate that MedAugment achieves the highest consistency between predictions and GT pairs. Specifically, it achieves the second-lowest 1.62\% mean and the lowest 2.39\% standard deviation. Although MonoAugment demonstrates the optimal mean of 1.56\%, MedAugment delivers comparable performance with only a marginal difference. However, the TriAugment does not exhibit superior performance, achieving the highest mean and standard deviation of 3.90\% and 4.75\%. Concerning point distribution, most points are located within the 95\% limits of agreement across methods, demonstrating strong consistency between the predicted and GT masks. An exception is the TriAugment, which presents more outliers beyond the agreement limits. Regarding trend differences, these methods do not exhibit noticeable systematic trend lines, suggesting the absence of consistent biases in the form of observable trends.

\subsection{Ablation Study}
\label{5.3}

\begin{figure}
    \centering
    \includegraphics[width=\linewidth]{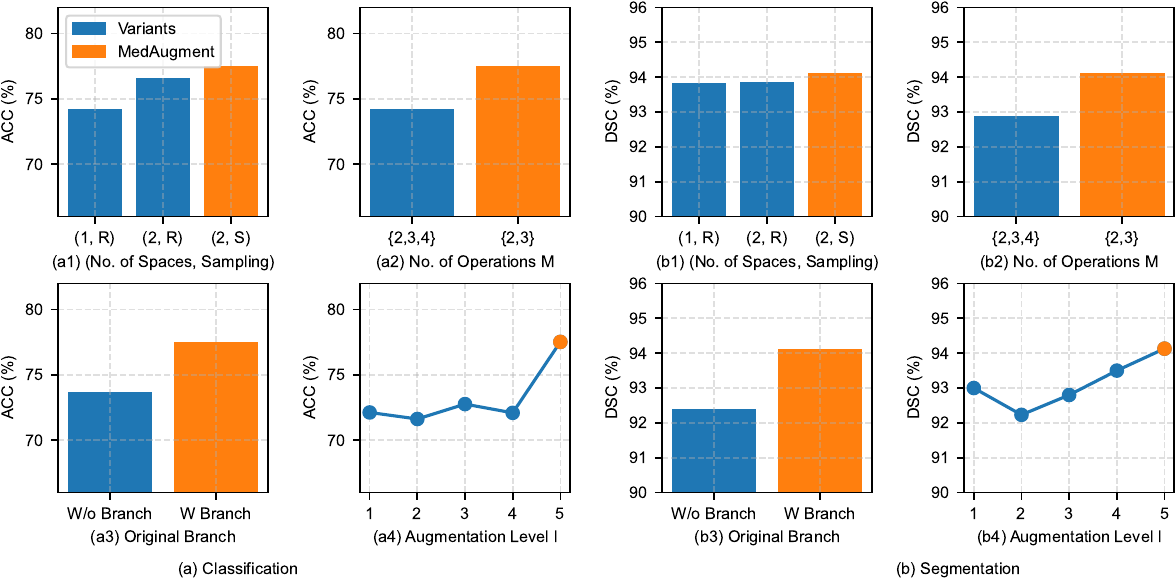}
    \caption{Ablation performance of MedAugment and its ablation variants on the LUNG dataset using ResNet and UNet. R denotes random sampling, and S stands for sampling using the proposed strategy.}
    \label{fig7}
\end{figure}

We perform ablation experiments to analyze the contribution of each component in MedAugment on the LUNG dataset using ResNet for classification and UNet for segmentation in \Cref{fig7}. We adopt the LUNG dataset that is consistent across classification and segmentation tasks, producing more convincing and comparable results. We organize the ablation study into four groups. The first group comprises two variants, including one using two augmentation spaces without the proposed sampling strategy to evaluate the contribution of the sampling strategy, and one using a single augmentation space without the sampling strategy to evaluate the contribution of the augmentation space design. The second group includes a variant with more sequential operations to analyze the effect of the number of operations. The third group includes a variant without the retained original branch to validate its effectiveness. The last group consists of variants with different augmentation levels to investigate the influence of augmentation magnitude. The results reveal that each design component and configurations of MedAugment contributes to the performance. For classification, removing any component leads to a performance drop, and variants with different augmentation levels consistently present lower performance than MedAugment. Among the augmentation level variants, the highest performance is achieved at $l = 3$. Similar observations are found for segmentation, where each component contributes to performance improvement, and the best performance among the level variants is achieved at $l = 4$.

\subsection{Generalization}
\label{5.4}

\begin{figure}
    \centering
    \includegraphics[width=\linewidth]{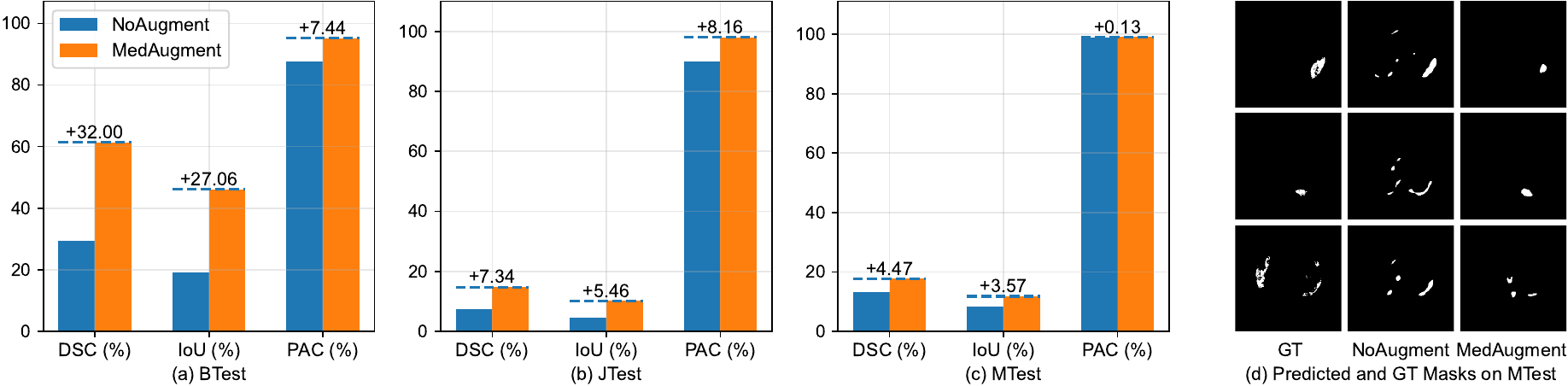}
    \caption{Generalization performance of MedAugment across different COVID components using UNet. BTest denotes training and validation on the J and M components and testing on the B component. The JTest and MTest are defined analogously.}
    \label{fig8}
\end{figure}

We conduct generalization experiments to validate the cross-domain generalization capability of MedAugment across different COVID components using UNet in \Cref{fig8}. We use the COVID dataset because it comprises three sources with varied data characteristics, which we denote as the B, J, and M components. We perform a three-fold evaluation, where the model is trained and validated on two components and tested on the remaining one. This results in three configurations, with the test subset being the B, J, and M components, respectively. Results demonstrate that MedAugment substantially enhances the generalization ability. Specifically, DSC improvements of 32.00\%, 7.34\%, and 4.47\% are observed after introducing MedAugment. Similar improvements are also reflected in the predicted masks, where MedAugment generates fewer erroneously predicted pixels and more accurate contour delineation. One observation is that MedAugment may not handle regions containing a large amount of fine debris well. In particular, it tends to miss these scattered small structures, suggesting that MedAugment may exhibit suboptimal performance for certain data patterns.

%% ------------------------------- Conclusions ------------------------------- %%

\section{Conclusions}
\label{6}

Here, we propose a suitable yet general automatic DA method termed MedAugment for MIA. We develop pixel and spatial augmentation spaces and exclude operations that can disrupt the details and features within medical images. This can prevent the involvement of severe color distortions or structural alterations to undermine the medical diagnostic value. Besides, we propose a sampling strategy by constraining the number of operations sampled from the proposed spaces. Furthermore, we formulate a hyperparameter mapping relationship to produce a rational augmentation level and ensure that the proposed approach is fully controllable with a single hyperparameter. These designs can address the differences between natural and medical images. Extensive experimental results on eight medical datasets demonstrate the effectiveness of MedAugment. Despite this, MedAugment may still face challenges in adapting to non-standard data characteristics. As it is currently designed for general medical imaging, it may exhibit suboptimal performance on certain modalities, object categories, or data patterns. A feasible improvement is to tailor the augmentation strategy to different data characteristics. This can be quantitatively achieved by leveraging characteristic-specific properties, such as object size, boundary complexity, or brightness level. For example, data containing smaller or scattered structures may require larger scaling factors. In addition, MedAugment is currently designed and evaluated for 2D medical images, and many target modalities, such as computed tomography and magnetic resonance imaging, are inherently volumetric. A volumetric extension is non-trivial and would require adapting the proposed spatial operations from 2D to 3D to preserve anatomical consistency across slices. It would also require enforcing coherent augmentation along the through-plane direction to avoid slice-wise inconsistencies.

%% ---------------------------------------- Others ---------------------------------------- %%

% \section*{Data availability statement}
% The related data of this research can be accessed through the author upon reasonable request.

%% ---------------------------------------- Reference ---------------------------------------- %%

%% If you have bibdatabase file and want bibtex to generate the
%% bibitems, please use

\bibliographystyle{unsrt}
\bibliography{reference.bib}

\biboptions{sort&compress}

%% else use the following coding to input the bibitems directly in the
%% TeX file.

% \begin{thebibliography}{00}

%% \bibitem[Author(year)]{label}
%% Text of bibliographic item

% \bibitem[ ()]{}

% \end{thebibliography}

%% ---------------------------------------- Biography ---------------------------------------- %%

% % \hspace*{\fill} % gap
% \subsection*{  }
% \setlength\intextsep{0pt} % align
% \begin{wrapfigure}{l}{30mm}
%     \centering
%     \includegraphics[width=1.2in,height=1.5in,clip,keepaspectratio]{liu.png}
% \end{wrapfigure}
% \noindent \textbf{Zhaoshan Liu} received his bachelor's degree in 2020 and his master's degree in 2021. He is currently pursuing his Ph.D. at the College of Design and Engineering, National University of Singapore. His research interests include computer vision and its applications in medical image analysis.\par

% % \hspace*{\fill} % gap
% \subsection*{  }
% \setlength\intextsep{0pt} % align
% \begin{wrapfigure}{l}{30mm}
%     \centering
%     \includegraphics[width=1.2in,height=1.5in,clip,keepaspectratio]{shen.png}
% \end{wrapfigure}
% \noindent \textbf{Lei Shen} is a Senior Lecturer in Mechanical Engineering, at the National University of Singapore. His interest lies in computational materials, physics, and chemistry based on DFT, high-throughput calculations, and machine learning. He has published 150+ papers with 7000 citations and an h-index of 47.\par

%% ---------------------------------------- End ---------------------------------------- %%

\end{sloppypar}
\end{document}